# Realization of 2D Crystalline Metal Nitrides via Selective Atomic Substitution

Jun Cao[1†], Tianshu Li[2†], Hongze Gao[2], Yuxuan Lin[4], Xingzhi Wang[1], Haozhe Wang[4], Tomás Palacios[4], Xi Ling[1,2,3]*

1. Department of Chemistry, Boston University, 590 Commonwealth Avenue, Boston, MA 02215, USA.

2. Division of Materials Science and Engineering, Boston University, 15 St. Marys Street, Boston, MA 02215, USA.

3. The Photonics Center, Boston University, 8 St. Marys Street, Boston, MA 02215, USA.

4. Department of Electrical Engineering and Computer Science, Massachusetts Institute of Technology, Cambridge, MA 02139, USA.

[†]These authors contributed equally to this work.

*Corresponding author. Email: xiling@bu.edu (X.L.)

**Abstract**

Two-dimensional (2D) transition metal nitrides (TMNs) are new members in the 2D materials family with a wide range of applications. Particularly, highly crystalline and large area thin films of TMNs are potentially promising for applications in electronic and optoelectronic devices; however, the synthesis of such TMNs has not yet been achieved. Here, we report the synthesis of few-nanometer thin $Mo_5N_6$ crystals with large area and high quality via *in situ* chemical conversion of layered $MoS_2$ crystals. The structure and quality of the ultrathin $Mo_5N_6$ crystal are confirmed using transmission electron microscopy, Raman spectroscopy and X-ray photoelectron spectroscopy. The large lateral dimensions of $Mo_5N_6$ crystals are inherited from the $MoS_2$ crystals that are used for the conversion. Atomic force microscopy characterization reveals the thickness of $Mo_5N_6$ crystals is reduced to about 1/3 of the $MoS_2$ crystal. Electrical measurements show the obtained $Mo_5N_6$ samples are metallic with high electrical conductivity (~ 100 $\Omega$ sq$^{-1}$), which is comparable to graphene. The versatility of this general approach is demonstrated by expanding the method to synthesize $W_5N_6$ and TiN. Our strategy offers a new direction for preparing 2D TMNs with desirable characteristics, opening a door for studying fundamental physics and facilitating the development of next generation electronics.

**Introduction**

Transition metal nitrides (TMNs) are well-known for their high melting points, hardness and chemical inertness (*1*, *2*). The successful preparation of two-dimensional (2D) MXenes through selectively etching the "A" layer of bulk MAX phase triggered the study of 2D metal carbides and nitrides since 2011 (*3–6*). Although a variety of carbide MXenes have been synthesized using this method, the synthesis of nitrides MXenes is limited to $Ti_4N_3$ and $Ti_2N$ (*7*, *8*), due to the low availability of bulk MAXs for corresponding 2D metal nitrides. Urbankowski et al. further reported the synthesis of multilayer molybdenum and vanadium nitrides by ammoniation of carbide MXenes synthesized using the selective etching method (*9*). Currently, as-synthesized 2D TMNs are either in solution phase or in powder form, and the lateral sizes of the flakes are usually small (< 10 micrometers) (*10–15*). In addition, samples prepared by selective etching suffer from defects generated during the etching process and large thickness distributions (*16*, *17*). These types of 2D TMNs are not suitable for high performance electronic and optoelectronic device applications which often require 2D crystals with decent area (>10 micrometers of lateral size), quality, and deposition on solid state substrates (*18*, *19*). Conventional 2D materials such as graphene and transition metal dichalcogenides (TMDs) produced using chemical vapor deposition (CVD) have made significant progress in advancing the development of electronic devices (*20–22*). In contrast to many conventional 2D materials with strong in-plane covalent bonds and weak out-of-plane van der Waals interactions between the layers, the van der Waals gap is missing in bulk TMN crystals. Instead, covalent bonding extends in three-dimensional (3D) frameworks of TMNs. Thus, conventional methods such as top-down mechanical exfoliation and bottom-up CVD, which take advantage of the weak interlayer van der Waals interaction and self-limited in-plane growth, are ineffective for the synthesis of large-area and ultrathin 2D TMNs. Therefore, developing effective routes to synthesize 2D TMN crystals with desired morphology and quality is essential to realize the potential for applications toward electronic devices.

Using nitriding reaction of metal oxides, TMDs and transition metal carbides (TMCs) to synthesize bulk TMNs is an established process (*23–28*). Recently, transformation of 2D $MoS_2$ and TMCs to 2D TMNs while retaining crystal structure was demonstrated in an ammonia atmosphere (*9*, *27*). This transformation method is promising for large-area synthesis of 2D TMNs, considering the well-established preparation methods TMDs thin films with controlled thickness and various lateral dimensions ranging from micrometer size to wafer scale (*22*, *29*, *30*). However, the precursors used in the reported work for the transformation are either multilayered carbide MXenes or powder TMDs (*9*, *27*). The obtained 2D TMNs are still marred by large thickness distributions, defective structures and small lateral sizes passed on from the precursors. Furthermore, an atomic level investigation of the transformation, which is the key to evaluate the potential of the strategy for producing high-quality and large-area 2D TMN crystals, is lacking.

Here, beginning with a precursor layer of $MoS_2$, an *in situ* transformation from $MoS_2$ to ultrathin $Mo_5N_6$ is achieved. Optical and scanning electron microscopic (SEM) images show that the geometry and morphology of the crystals are retained during the transformation. The obtained $Mo_5N_6$ exhibits high crystallinity over the entire area. *In situ* transformations of various $MoS_2$ flakes down to 4 layers for $Mo_5N_6$ flakes with different thicknesses are achieved. Atomic force microscopic (AFM) analysis shows that the thickness of most of $MoS_2$ flakes is reduced by about 2/3 after the transformation. Electrical measurements show the high conductivity of metallic $Mo_5N_6$ samples. We further demonstrate that such strategy can be applied to other TMDs such as $WS_2$ and $TiS_2$ for preparing their corresponding TMNs. This work opens a new direction for preparing 2D TMNs with desired quality that are previously inaccessible.

**Results**

Figure 1 shows a schematic illustration of the typical reaction process. The conversion was performed in a horizontal tube furnace (See Materials and Methods for details of synthesis). $MoS_2$ flakes with different thicknesses were prepared using mechanical exfoliation and transferred onto a $SiO_2$/Si substrate, which was then placed at the center of the furnace. A crucible filled with urea was placed in the upstream zone of the furnace, serving as the ammonia gas source which was released from the thermolysis of urea at 200 °C (*31*, *32*). 100 sccm Ar gas was used as carrier gas. The temperature in the center of the furnace was set at 750~800 °C for the chemical reaction between gaseous ammonia and $MoS_2$ flakes (*23*, *27*). Apart from $Mo_5N_6$ which was the solid product, the other two side products, $H_2S$ and $N_2$ were in gaseous phase and carried away by the Ar gas (*23*, *27*), leaving clean $Mo_5N_6$ products with minimum contaminants.

Figure 2 (A and B) shows optical images of a 12.9 nm $MoS_2$ flake before and after the reaction. The optical contrast of the flake changed dramatically after the reaction, but the morphology and shape of flakes are retained. Conversion results of more $MoS_2$ flakes with different thicknesses ranging from ~3 nm to tens of nanometers are shown in fig. S1. All $MoS_2$ flakes displayed significant changes of optical contrast after conversion. Absorption spectra (fig. S2) of the flake before ($MoS_2$) and after ($Mo_5N_6$) the conversion suggest that such change originated from the distinct optical properties, consistent with the distinct electronic band structures of the two materials (*15*, *33*). More evidence supporting the successful conversion of $MoS_2$ to $Mo_5N_6$ is reported in the later section. Atomic force microscopy (AFM) images show the atomically smooth surface of the flakes after conversion (fig. S3), suggesting a mild conversion process during which the reaction is confined within the original $MoS_2$ flakes. We further measured the Raman and photoluminescence (PL) spectra of the flake before and after the chemical transformation. As shown in Fig. 2C, pronounced $A_{1g}$ and $E^1_{2g}$ modes at ~384 and ~407 cm$^{-1}$ (*34*, *35*) were observed from MoS$_2$ flakes before the transformation. Both modes vanished completely after the reaction and new Raman peaks at ~215 and 710 cm$^{-1}$ appeared, indicating the transformation of the crystal domain. Similarly, PL from MoS$_2$ disappeared after the conversion (Fig. 2D) (*33*, *36*), matching well with the semiconducting and metallic property of MoS$_2$ and Mo$_5$N$_6$ (*15*, *33*). In addition, Raman mapping results (Fig. 2, E and F) indicated the entire flake was converted from MoS$_2$ to Mo$_5$N$_6$, where no Raman signals from MoS$_2$ was observed on the converted flake globally. Corresponding PL mapping results are shown in fig. S4.

To determine the phase and crystal structure of the flake, we performed transmission electron microscopy (TEM) characterization on an as-prepared sample. The low magnification TEM image in Fig. 3A shows the smooth surface of the flake where the small variations of the contrast are due to the wrinkles generated during the transfer process. The selected-area electron diffraction (SAED) pattern in Fig. 3B (see also fig. S5) indicates the high crystallinity of the sample. Same SAED pattern was observed when measuring on different regions of the flake. The hexagonal diffraction pattern is consistent with the crystal structure of the Mo$_5$N$_6$ (*15*). The experimental and simulated SAED patterns (fig. S5) of Mo$_5$N$_6$ match well with each other. High-angle annular dark-field (HAADF) scanning transmission electron microscope (STEM) image is shown in Fig. 3C, where Mo atoms and atomic lattice are clearly seen and no obvious defect is found in the sample. N atoms are invisible under our STEM condition due to the small atomic number compared to Mo atoms. Additionally, the distance of 0.239 nm between the lattice planes is consistent with the d-spacing of (110) planes in Mo$_5$N$_6$ crystal, which matches with the SAED pattern. We also noticed that the obtained Mo$_5$N$_6$ crystal had an enhanced stability under 200 keV electron beam compared to MoS$_2$ flakes (*37*, *38*), where Mo$_5$N$_6$ flakes on TEM grid remained stable without noticeable change after being exposed under electron beam for 30 minutes (fig. S6), providing an additional advantage for future applications in electronic devices. STEM imaging and energy-dispersive spectroscopy (EDS) mapping were performed (see fig. S5) on a thick $Mo_5N_6$ flake which provides better signal-to-noise ratio of EDS signal. Dark field STEM image shows the inhomogeneous contrast of the thick $Mo_5N_6$ flake (fig. S5), which is probably due to strong local strain during the conversion process. EDS mapping of N K peak and Mo L peak (fig. S5) shows uniform distribution of Mo and N elements in the flake. The TEM characterization confirmed that the highly crystalline $Mo_5N_6$ flake was successfully obtained through the chemical transformation.

The chemical composition and oxidation states of the elements in sample $Mo_5N_6$ were characterized using X-ray photoelectron spectroscopy (XPS) and EDS analyses. Figure 3D shows the XPS wide scan survey spectrum of the sample, where significant Mo and N signals are observed and no signal from S appeared, indicating complete conversion of $MoS_2$ to $Mo_5N_6$. Other elements such as O, Si and C are from the $SiO_2$/Si substrate and the chamber environment. Similar results were obtained from EDS characterization (fig. S7). High-resolution XPS spectra deconvolution for the N 1s, Mo 3p and Mo 3d regions are shown in Fig. 3 (E and F). The N 1s peak at 397.9 eV and Mo $3p_{3/2}$ at 395.3 eV (Fig. 3E) suggest that chemical bond formed between Mo and N (*15*). In the Mo 3d region (Fig. 3F), the peaks at 233.0 eV and 229.8 eV were assigned to the binding energies of Mo $3d_{3/2}$ and Mo $3d_{5/2}$, respectively (*39*), with a spin-orbit splitting of 3.2 eV. This characteristic doublet of core-level Mo $3d_{5/2}$ and $3d_{3/2}$ indicates that Mo (+4) oxidation state dominates in $Mo_5N_6$ (*25*). Nevertheless, the binding energy of 229.8 eV is slightly smaller than reported value at 230.0 eV for Mo (+4) oxidation state (*15*), suggesting the co-existence of Mo (+3) oxidation state in the sample. No peaks corresponding to higher oxidation states of Mo from $MoO_x$ were observed in the Mo 3d region (*40*), ruling out the possibility of the formation of $MoO_x$.

To address the mechanism by which the lattice changed during the atomic substitution from $MoS_2$ to $Mo_5N_6$, we investigated thickness changes of flakes before and after chemical conversion. The crystal structure models of $MoS_2$ and $Mo_5N_6$ (Fig. 4, A and B) predict that the van der Waals gap between $MoS_2$ layers will disappear during the transformation to $Mo_5N_6$ crystals, leading to the reducing of the thickness from $MoS_2$ to $Mo_5N_6$. The distance between two Mo layers in $MoS_2$ is 7.66 Å (*41*), while it is 2.77 Å in $Mo_5N_6$ (*42*), therefore the thickness of the flake can be predicted to decrease to 36% when transforming from $MoS_2$ to $Mo_5N_6$. To test this hypothesis, we performed AFM to extract the thickness of flakes. As expected we observed that the thickness was reduced post chemical transformation. Figure 4C and D show typical AFM images of $MoS_2$ and corresponding $Mo_5N_6$ flake, where they have nearly the same morphology and both surfaces are smooth. However, the thickness of the flake decreased from 9.5 nm for $MoS_2$ to 3.5 nm for $Mo_5N_6$ (Fig. 4E), where $Mo_5N_6$/$MoS_2$ thickness ratio is 37%. Furthermore, we performed this comparison for ~30 flakes with $MoS_2$ thickness ranging from few nanometers to ~35 nm. Reduced sample thickness was observed for all flakes after conversion (Fig. 4F). The average $Mo_5N_6$/$MoS_2$ thickness ratio is 56% based on the slope of linear fitting curve, but for $MoS_2$ flakes with the thickness of 5 to15 nm, the ratio is around 40%, matching well with the expected value. For thinner flakes, the ratio is larger, probably because the termination groups on the surface are non-negligible at this size range (*3*, *5*) or the substrate roughness introduces a large uncertainty for the measurement. Figure S8 shows a 4-layer $MoS_2$ flake (3.4 nm) turned to a 2.1 nm $Mo_5N_6$ flake after conversion. Note that we found the morphology and quality of the flake remain well at such thin level. Based on the morphology retaining and thickness depression phenomena observed in our experiments, we propose the following mechanism of the chemical transformation from $MoS_2$ to $Mo_5N_6$. When $NH_3$ gas diffuses into the van der Waals gaps of the $MoS_2$ flake, one N atom will replace two S atoms sandwiched by two adjacent Mo layers. When Mo-S bonds around a Mo atom are broken, six binding sites on the Mo atom will be released. Each N atom will then bond with six Mo atoms to form $Mo_5N_6$, where adjacent Mo layers are bridged by the N atoms, leading to the vanishing of the van der Waals gaps that originally exist in $MoS_2$. In this process, breaking Mo-S bonds and forming Mo-N bonds do not require large rearrangements of Mo atoms, even though small shift of positions of Mo atoms may be necessary to compensate the bond length difference between Mo-S and Mo-N. In fact, despite the layered structure of $MoS_2$ (point group: $D_{6h}$, space group: $P6_3/mmc$) (*41*) and non-layered nature of $Mo_5N_6$ (point group: $C_{6h}$, space group: $P 6_3/m$) (*42*), the Mo layers in these two materials are both in hexagonal pattern. This similarity in structure allows gentle transformation without structural collapse, leading to the morphology retaining of samples before and after chemical transformation.

To further investigate the transformation process, we performed the reaction under varying conditions. We found that excess amount of urea, which provided sufficient $NH_3$ gas in the chamber, was a key factor for the success because it ensured a reducing environment to prevent $MoS_2$ from oxidation (fig. S9). Additionally, the degree of conversion could be tuned by changing the reaction temperature and time. A complete conversion occurred within 5 minutes when the reaction temperature varied from 750 °C to 800 °C (fig. S9). However, at 700 °C, the conversion was only partially complete even after 60 minutes, where only the edge area was converted and no $Mo_5N_6$ was observed in the central region of the flake (fig. S10), indicating that the conversion started from the edge of the $MoS_2$ flake. Because of the partial conversion, we realized a lateral heterostructure between $MoS_2$ and $Mo_5N_6$ through a simple partial conversion. The successful synthesis of such heterostructure offers a great platform for future study on the junction properties and applications. Moreover, the converted $Mo_5N_6$ showed excellent stability, where the crystal structure remained intact after 6 months and survived in acetone, DI water and 1 mol/L $H_2SO_4$ solution for at least 2 hours without noticeable changes from optical images and Raman spectra (fig. S11). This is advantageous to be compatible with future device fabrication process.

To examine the property of the converted $Mo_5N_6$ sample, we fabricated a back gate transport device on a 9.5 nm thick $Mo_5N_6$ flake (Fig. 5A and fig. S13). The gate-dependence of the transport current shows that the drain-source current ($I_{ds}$) remained constant as the gate voltage scanned from -20 V to 20 V and variations of $I_{ds}$ was still negligible even when noise signal became visible, featuring metallic behavior of the as-synthesized $Mo_5N_6$ sample (Fig. 5B). We further performed the temperature-dependent transport measurement down to 77 K to investigate the electrical conductivity of the sample. Fig. 5C shows the output *I-V* characteristics through a 4-probe measurement where slope of the *I-V* curves increases very slowly as the temperature increases from 77 K to 240 K (Fig. 5C, inset). The sheet resistances of $Mo_5N_6$ at different temperatures were extracted from the slope of *I-V* curves and the dimension of the transport device (Fig. 5D), using the formula $R_s$ = R × (W/L). R is the total electrical resistance of the device, and W and L are the effective width and length of the measurement area, which are measured as 3.4 and 6.1 µm from the optical image, respectively. Low sheet resistances of $Mo_5N_6$ ranging from 114.4 to 117.8 Ω sq$^{-1}$ were obtained under different temperatures, which is in the same order of magnitude of CVD graphene (*43*). The fact that the sheet resistance of $Mo_5N_6$ is not sensitive to the temperature change from 77 K and 240 K is consistent with the trend of transition metal nitrides reported in the literature (*9*).

We further applied this strategy to the conversion from $WS_2$ and $TiS_2$ to their corresponding nitrides (See Materials and Methods for details of the synthetic conditions). Figure 6 shows a typical conversion result on a 5.6 nm $WS_2$ and a ~100 nm $TiS_2$ flake. As shown in Fig. 6A-H, significant optical contrast change was observed from $WS_2$ and $W_5N_6$, as well as from $TiS_2$ to TiN. Similar to the case of $MoS_2$ to $Mo_5N_6$ conversion, the morphology of the flakes was retained. Raman characterization showed that the Raman signatures from $WS_2$ and $TiS_2$ disappeared completely (*44, 45*). Instead, new peaks that correspond to the phonon modes of $W_5N_6$ (e.g. ~258 cm$^{-1}$) and TiN (e.g. ~154 and 620 cm$^{-1}$) (*46, 47*) appeared (Fig. 6, I to J), indicating a successful chemical conversion. The Raman mapping results clearly show that the conversion is thorough and uniform. More optical and corresponding SEM images of $W_5N_6$ (fig. S12) and TiN (fig. S12) corroborate uniform surface of converted samples. Phase and crystallinity of $W_5N_6$ are characterized by HRTEM and EDS analyses (fig. S14) (*48*). Similar to $Mo_5N_6$, $W_5N_6$ and TiN also exhibit excellent stability against acetone, DI water and 1 mol/L $H_2SO_4$ for at least 2 hours (fig. S11).

**Discussion**

We demonstrate a versatile conversion strategy from layered TMDs to their corresponding ultrathin nitrides through atomic substitution from chalcogen to nitrogen. This method facilitates the production of 2D crystalline TMNs including $Mo_5N_6$, $W_5N_6$ and TiN, offering a pathway toward an important class of 2D materials that previously are inaccessible. The investigation of the *in situ* transformation from $MoS_2$ layers to 2D $Mo_5N_6$ crystals shows reduced thickness after conversion. 2D $Mo_5N_6$ crystal as thin as 2.1 nm is achieved by converting a 4-layer $MoS_2$. The electrical measurement shows that the converted $Mo_5N_6$ is metallic with high conductivity of about 100 Ω sq$^{-1}$. By controlling the reaction rate, we achieve a $Mo_5N_6$-$MoS_2$ lateral heterostructure, demonstrating the advantage of our method in integrating 2D materials together for basic building blocks (e.g. metal-semiconductor junction) for future applications for electronic devices. More importantly, building on the success of the TMDs synthesis in the field (*21*, *22*, *30*), we anticipate the strategy we report here will lead to effective synthesis of large-area and high quality 2D TMNs that satisfy the needs for high performance electronic and optoelectronic devices.

## MATERIALS AND METHODS

### Conversion from TMDs to TMNs

$MoS_2$, $WS_2$ and $TiS_2$ with different thicknesses were prepared using mechanical exfoliation method from their bulk crystals (purchased from HQ Graphene) and transferred onto 300 nm $SiO_2$/Si substrates. Note that $SiO_2$/Si substrates were cleaned by sonication in water, acetone, and isopropanol solvents sequentially (each 10 minutes), followed by $O_2$ plasma cleaning before use. Chemical conversions were conducted in a horizontal tube furnace where Ar gas was introduced into the tube and the flow was controlled by a mass flow controller. Samples were placed in a 1-inch diameter quartz tube in the center of the furnace. Excess amount of urea (500 mg) was placed in an $Al_2O_3$ crucible located upstream of the furnace, serving as nitrogen source. After purging with Ar gas for 10 minutes, the tube was heated to 800 ℃ with a 30-min ramp. The conversion time for $Mo_5N_6$ and TiN is 1 hour, and it is 2 hours for $W_5N_6$. Throughout the conversion, 100 sccm Ar flow was used to maintain the inert atmosphere. When conversion was completed, the system was cooled down naturally.

### Materials characterizations

As-prepared TMNs were characterized using Raman spectroscopy, AFM, TEM, STEM, SAED XPS and EDS. Raman and PL measurements were performed on a Renishaw inVia Raman microscope equipped with a 532 nm laser line. All the spectra in comparison were taken using the same condition. The AFM topography was acquired on a Bruker Dimension system. TEM measurements were performed on a FEI Tecnai Osiris transmission electron microscope, operating at a 200 keV accelerating voltage. SAED were measured on a JEOL 2100 transmission electron microscope. The SAED simulation was perform through STEM_CELL software (*49*). Atomic resolution STEM and EDS mapping was performed on a JEOL ARM 200F scanning transmission electron microscope, operating at 200 keV. XPS measurements were performed under ultrahigh vacuum below $2\times10^{-9}$ Torr using monochromatic Al Kα radiation at 1486.7 eV on a Surface Science Instruments SSX100 spectrometer.

**Fabrication and conductivity measurements**

The electrical transport device was fabricated through laser writer lithography and thermal evaporation of 5 nm Cr and 45 nm Au. For lift-off, the sample was soaked in Remover PG at 60°C for 10 minutes to remove the photoresist and then washed in isopropanol and deionized water. Device characterization was performed using a semiconductor parameter analyzer (Keysight B1500A) and a Lakeshore cryogenic probe station with micromanipulation probes and liquid nitrogen cooling. All measurements were done in vacuum ($<3\times10^{-6}$ Torr).

**SUPPLEMENTARY MATERIALS**

**Fig. S1. Typical optical images of MoS$_2$ and Mo$_5$N$_6$ flakes with different thicknesses.**

**Fig. S2. Absorption spectra of MoS$_2$ and Mo$_5$N$_6$ samples on quartz substrate.**

**Fig. S3. AFM images of MoS$_2$ and Mo$_5$N$_6$ flakes in Figure (2).**

**Fig. S4. Maps of PL intensity at 673 nm of MoS$_2$ (A) and Mo$_5$N$_6$ (B) flakes in Figure 2.**

**Fig. S5. Structural and elemental characterizations of Mo$_5$N$_6$.**

**Fig. S6. TEM images of Mo$_5$N$_6$ sample under 200 keV electron beam.**

**Fig. S7. EDS spectrum of Mo$_5$N$_6$.**

**Fig. S8.** Optical, AFM and SEM images of chemical transformation on a 4-layer $MoS_2$ flake.

**Fig. S9.** Optical images of $Mo_5N_6$ flakes prepared from chemical transformations at different conditions.

**Fig. S10.** Optical images and Raman spectra of a partially converted flake at 700 °C.

**Fig. S11.** Stability test of $Mo_5N_6$, $W_5N_6$ and TiN.

**Fig. S12.** Optical and SEM images of $WS_2$, $W_5N_6$, $TiS_2$ and TiN flakes.

**Fig. S13.** Optical and AFM image of $Mo_5N_6$ transport device

**Fig. S14.** TEM and EDS characterizations of $W_5N_6$ converted from $WS_2$.

**Acknowledgments:** X. L. acknowledges the support of Boston University and Boston University Photonics Center. The device fabrication was done in Professor Kenneth Burch's lab at Boston College. Electrical measurements were conducted at MIT. Y. L. and T. P. acknowledge the partial support by the U.S. Army Research Office through the Institute for Soldier Nanotechnologies, under Cooperative Agreement number W911NF-18-2-0048, AFOSR FATE MURI, grant no. FA9550-15-1-0514, and the STC Center for Integrated Quantum Materials, NSF grant no. DMR 1231319. The TEM imaging was performed at the Center for Nanoscale Systems (CNS), a member of the National Nanotechnology Coordinated Infrastructure Network (NNCI), which is supported by the National Science Foundation under NSF award no. 1541959. CNS is part of Harvard University. **Author Contributions:** J.C., T.L. and X.L. conceived and designed the experiments. J.C. and H.G. performed the conversion, optical characterization and Raman measurements. T.L. performed TEM, STEM and AFM measurements. T.L. fabricated device. T.L. X.W. and Y.L. conducted electrical measurements supervised by X.L. and T.P. H.W. conducted XPS measurements. J.C., T.L. and X.L. analyzed the data and wrote the manuscript together with input from all authors. **Competing interests:** The authors declare that they have no competing interests. **Data and materials availability:** All data supporting the stated conclusions of the manuscript are in the paper and/or in the Supplementary Materials. Additional data related to this paper may be requested from the authors.

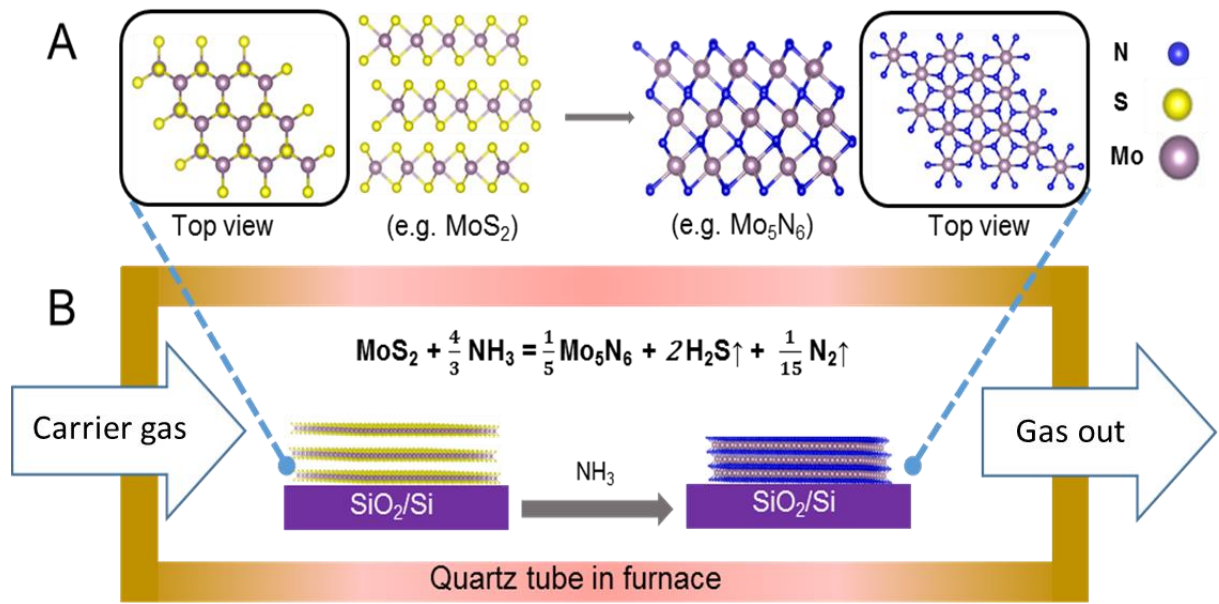

**Fig. 1. Schematic illustration of chemical conversion from MoS$_2$ to Mo$_5$N$_6$.** (**A**) Structural changes from MoS$_2$ to Mo$_5$N$_6$. (**B**) Schematics of the experimental setup for chemical conversion reactions.

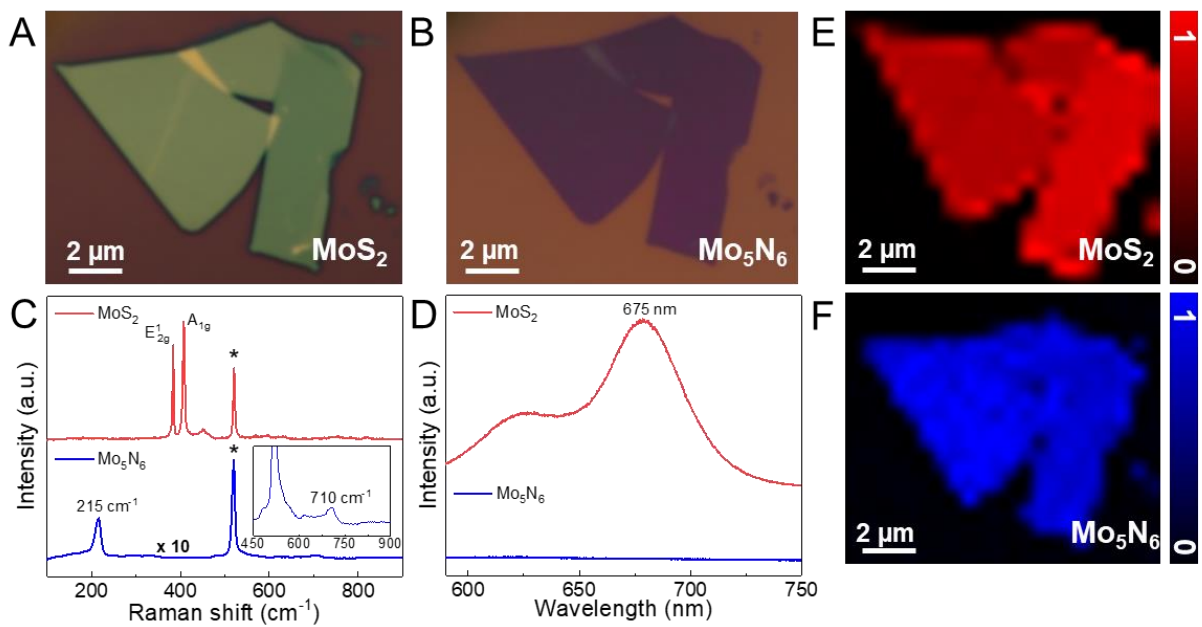

**Fig. 2. Typical optical microscope, Raman spectroscopy and PL characterization of $MoS_2$ and $Mo_5N_6$ flakes.** (**A**) Optical images of $MoS_2$ exfoliated on $SiO_2/Si$ substrate. (**B**) Optical image of $Mo_5N_6$ converted from $MoS_2$ in (**A**). (**C**) Comparison of Raman spectra of $MoS_2$ and $Mo_5N_6$. Peaks labeled with "*" are from the $SiO_2/Si$ substrate. (**D**) Comparison of PL spectra of $MoS_2$ and $Mo_5N_6$. (**E** and **F**) Raman intensity maps of the $A_{1g}$ mode of $MoS_2$ (**E**) and the 215 cm$^{-1}$ mode of $Mo_5N_6$ (**F**), respectively. Color bars show normalized Raman intensities, where "1" represent and "0" represent maximum and minimum intensity of Raman modes, respectively.

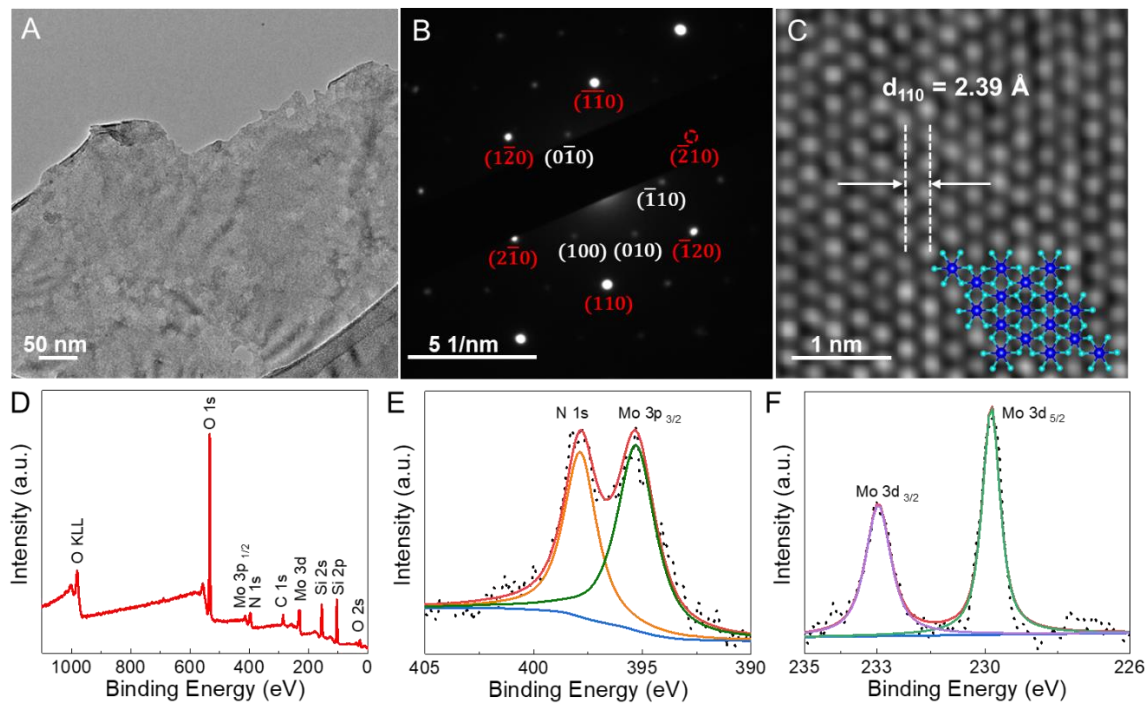

**Fig. 3. Crystal structure and elemental analysis of Mo$_5$N$_6$ samples converted from MoS$_2$.** (**A**) Low magnification TEM image of Mo$_5$N$_6$. (**B**) SAED pattern taken by a 25 cm camera. Diffraction planes are labelled according to SAED simulation in fig. S4. Red circle corresponds to a diffraction plane blocked by TEM beam stop. (**C**) Filtered HAADF STEM image. Hexagonal Mo pattern is clearly observed. Mo atoms and N atoms are labelled in blue and cyan color, respectively. (**D**) XPS survey spectrum of Mo$_5$N$_6$. (**E** and **F**) XPS spectra of Mo$_5$N$_6$ in Mo 3p and N 1s region (**E**) and Mo 3d region (**F**).

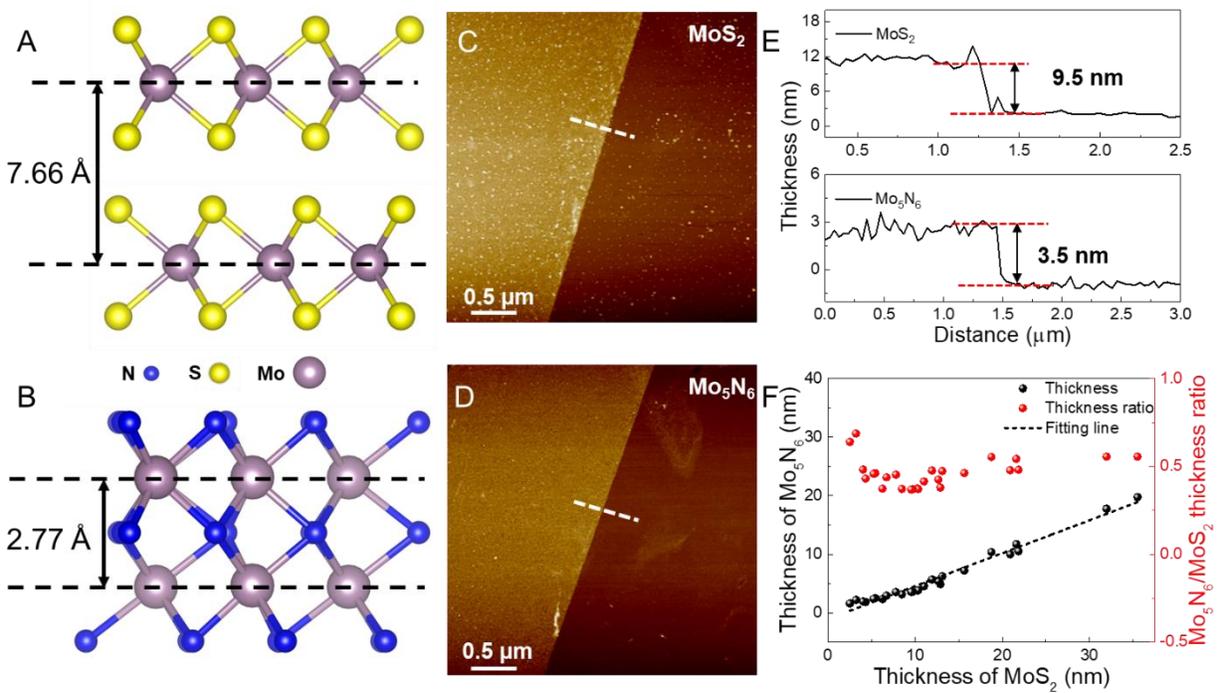

**Fig. 4. Thickness characterization and analysis of samples before (MoS₂) and after conversion (Mo₅N₆).** (**A** and **B**) Side view of crystal structures labeled with distance between adjacent Mo layers in MoS$_2$ and Mo$_5$N$_6$. (**C** and **D**) AFM images of the same flake before (MoS$_2$, **C**) and after conversion (Mo$_5$N$_6$, **D**). The white dashed lines indicate the location where we measured the thickness of the flakes. (**E**) AFM height profiles of the flake before (MoS$_2$, **C**) and after conversion (Mo$_5$N$_6$, **D**). (**F**) Correlation plot of the thickness of MoS$_2$ and converted Mo$_5$N$_6$ (in black). Data points in red show the corresponding thickness ratios of Mo$_5$N$_6$/MoS$_2$.

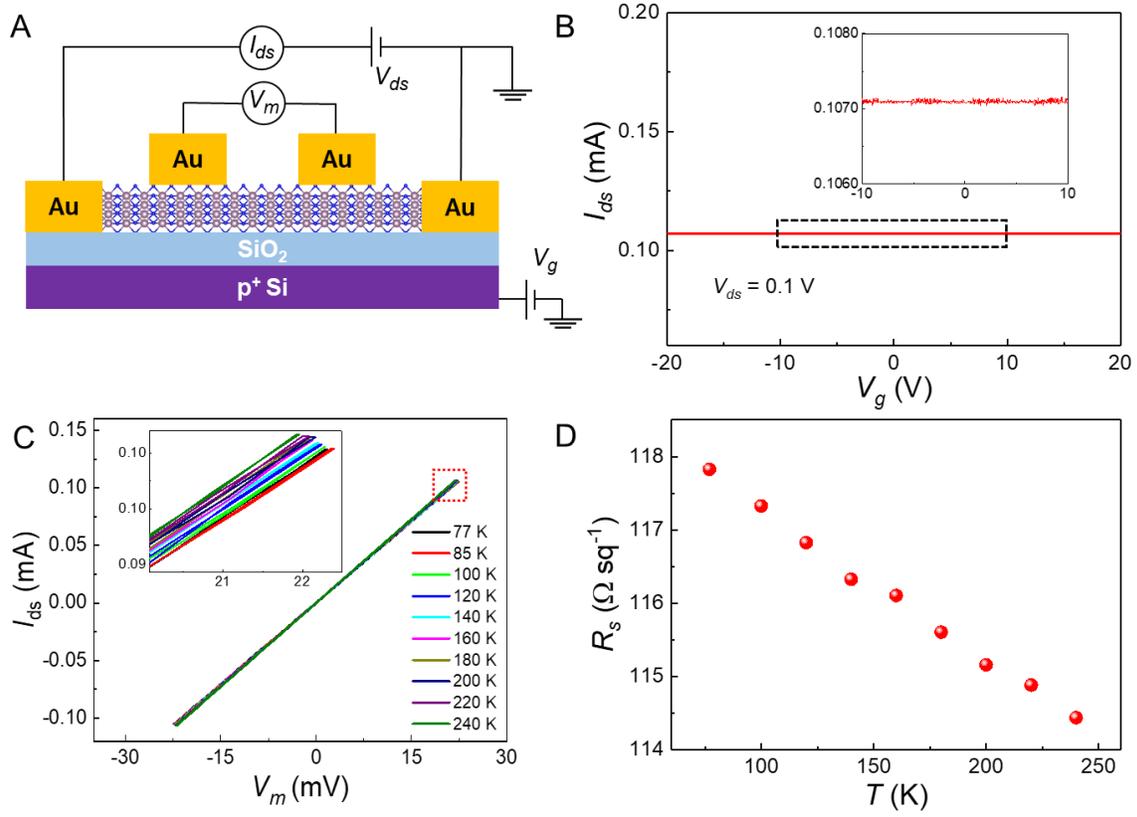

**Fig. 5. Electrical transport measurements of Mo$_5$N$_6$.** (**A**) Schematics of a back gate device together with electrical connections. (**B**) Transfer curve of Mo$_5$N$_6$ transport device for both forward and reverse $V_g$ bias with back gate modulations. Inset shows a zoom in image of the area indicated by the black rectangle. Negligible gate dependence of the $I_{ds}$ is observed in Mo$_5$N$_6$ transport device. (**C**) Output *I-V* curve of Mo$_5$N$_6$ transport device under different temperatures at zero gate voltage. Inset shows a zoom in region of the *I-V* curve indicated by the red square. (**D**) Temperature dependence of the sheet resistance of the Mo$_5$N$_6$ sample.

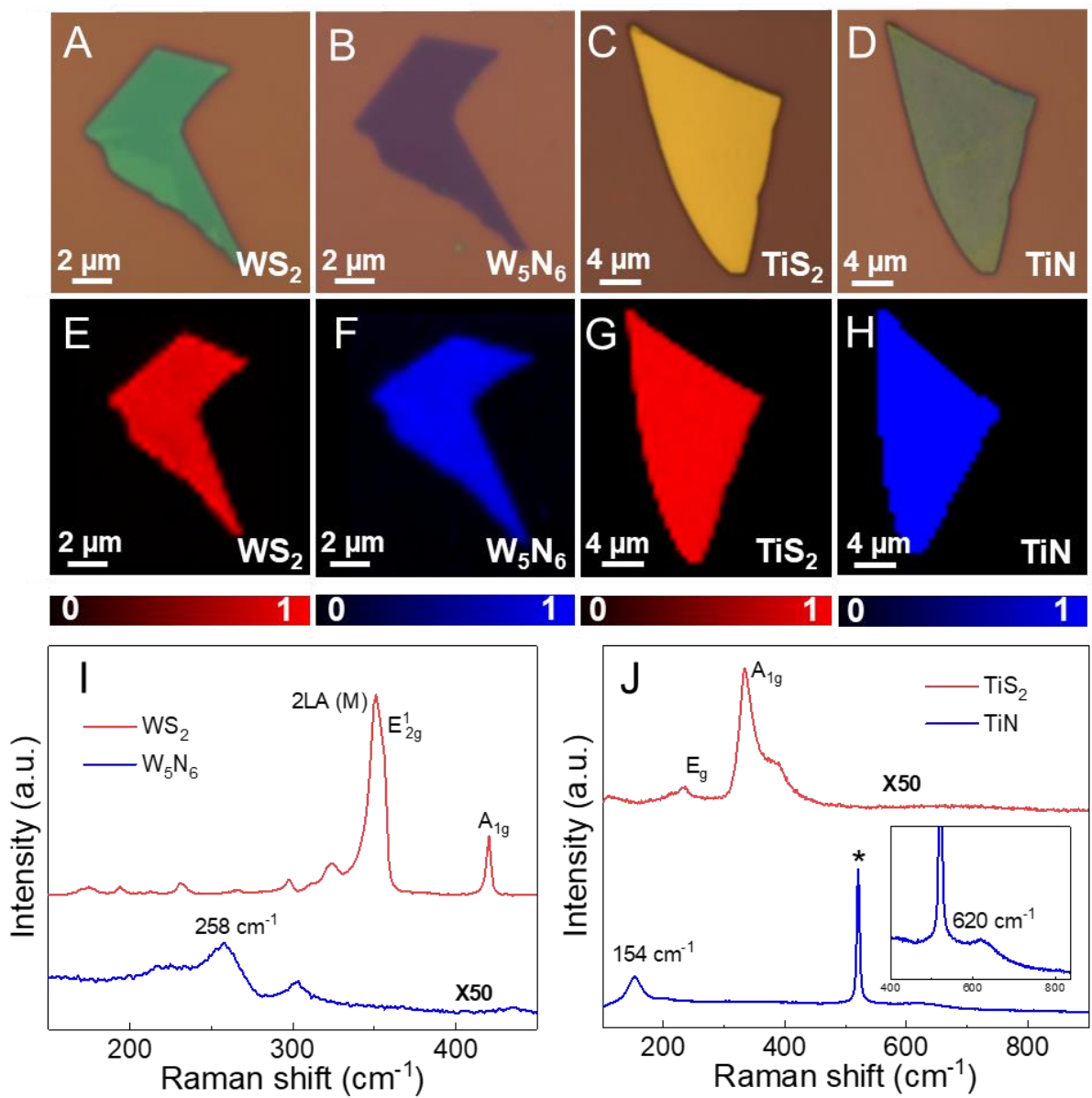

**Fig. 6. Conversion on WS₂ and TiS₂ for W₅N₆ and TiN.** (**A** to **D**) Optical images of WS$_2$ (**A**), W$_5$N$_6$ (**B**), TiS$_2$ (**C**), and TiN (**D**). (**E** to **H**) Raman intensity maps of E$^1_{2g}$ mode of WS$_2$ (**E**), 258 cm$^{-1}$ mode of W$_5$N$_6$ (**F**), A$_{1g}$ mode of TiS$_2$ (**G**) and 154 cm$^{-1}$ mode of TiN (**H**). Color bars show normalized Raman intensities, where "1" and "0" represent maximum and minimum intensity of Raman modes, respectively. (**I** and **J**) Comparison of Raman spectra of WS$_2$ and W$_5$N$_6$ (**I**), TiS$_2$ and TiN (**J**). Peak labeled with "*" is from the SiO$_2$/Si substrate.

# Supplementary Materials for

# Realization of 2D Crystalline Metal Nitrides via Selective Atomic Substitution

Jun Cao[1†], Tianshu Li[2†], Hongze Gao[2], Yuxuan Lin[4], Xingzhi Wang[1], Haozhe Wang[4], Tomás Palacios[4], Xi Ling[1,2,3]*

1. Department of Chemistry, Boston University, 590 Commonwealth Avenue, Boston, MA 02215, USA.

2. Division of Materials Science and Engineering, Boston University, 15 St. Marys Street, Boston, MA 02215, USA.

3. The Photonics Center, Boston University, 8 St. Marys Street, Boston, MA 02215, USA.

4. Department of Electrical Engineering and Computer Science, Massachusetts Institute of Technology, Cambridge, MA 02139, USA.

[†]These authors contributed equally to this work.

*Corresponding author. Email: xiling@bu.edu (X.L.)

This PDF file includes:

**Fig. S1.** Typical optical images of $MoS_2$ and $Mo_5N_6$ flakes with different thicknesses.

**Fig. S2.** Absorption spectra of $MoS_2$ and $Mo_5N_6$ samples on quartz substrate.

**Fig. S3.** AFM images of $MoS_2$ and $Mo_5N_6$ flakes in Figure (2).

**Fig. S4.** Maps of PL intensity at 673 nm of $MoS_2$ (A) and $Mo_5N_6$ (B) flakes in Figure 2.

**Fig. S5.** Structural and elemental characterizations of $Mo_5N_6$.

**Fig. S6.** TEM images of $Mo_5N_6$ sample under 200 keV electron beam.

**Fig. S7.** EDS spectrum of $Mo_5N_6$.

**Fig. S8.** Optical, AFM and SEM images of chemical transformation on a 4-layer MoS$_2$ flake.

**Fig. S9.** Optical images of Mo$_5$N$_6$ flakes prepared from chemical transformations at different conditions.

**Fig. S10.** Optical images and Raman spectra of a partially converted flake at 700 °C.

**Fig. S11.** Stability test of Mo$_5$N$_6$, W$_5$N$_6$ and TiN.

**Fig. S12.** Optical and SEM images of WS$_2$, W$_5$N$_6$, TiS$_2$ and TiN flakes.

**Fig. S13.** Optical and AFM image of Mo$_5$N$_6$ transport device

**Fig. S14.** TEM and EDS characterizations of W$_5$N$_6$ converted from WS$_2$.

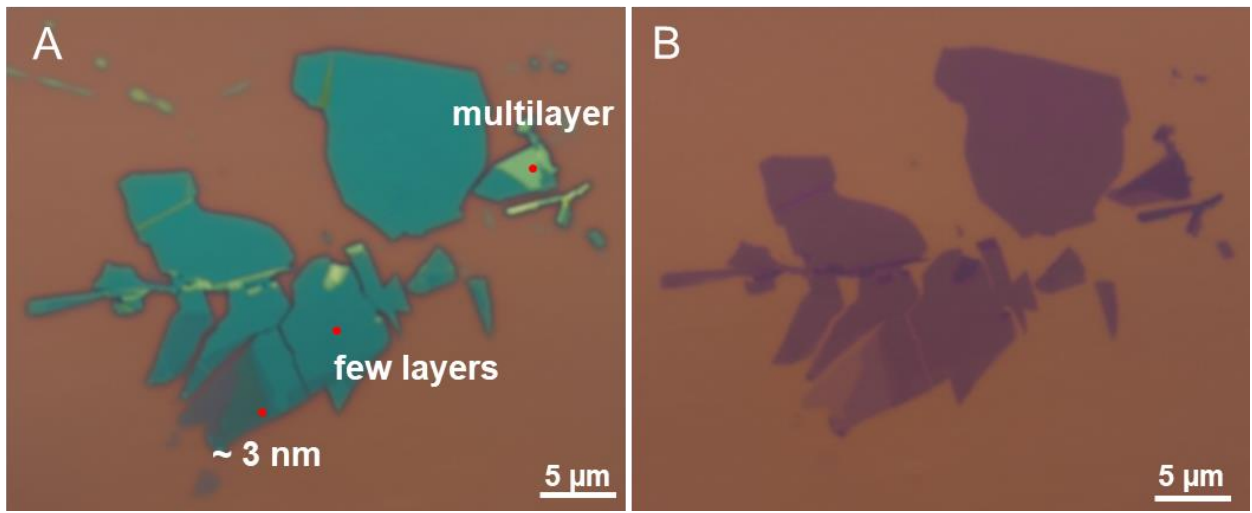

**Fig. S1. Typical optical images of MoS$_2$ and Mo$_5$N$_6$ flakes with different thicknesses. (A)** MoS$_2$ flakes before the conversion. **(B)** Mo$_5$N$_6$ flakes converted from MoS$_2$ flakes in **(A)**.

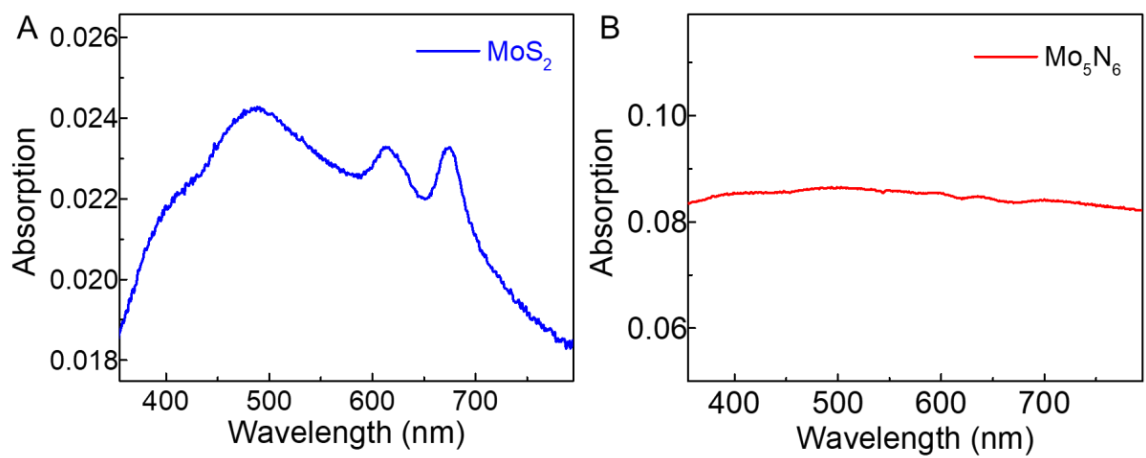

**Fig. S2. Absorption spectra of MoS$_2$ and Mo$_5$N$_6$ samples on a quartz substrate.** **(A)** Absorption spectrum of MoS$_2$ on a quartz substrate. **(B)** Absorption spectrum of Mo$_5$N$_6$ samples converted from MoS$_2$ flakes in **(A)** on the same quartz substrate.

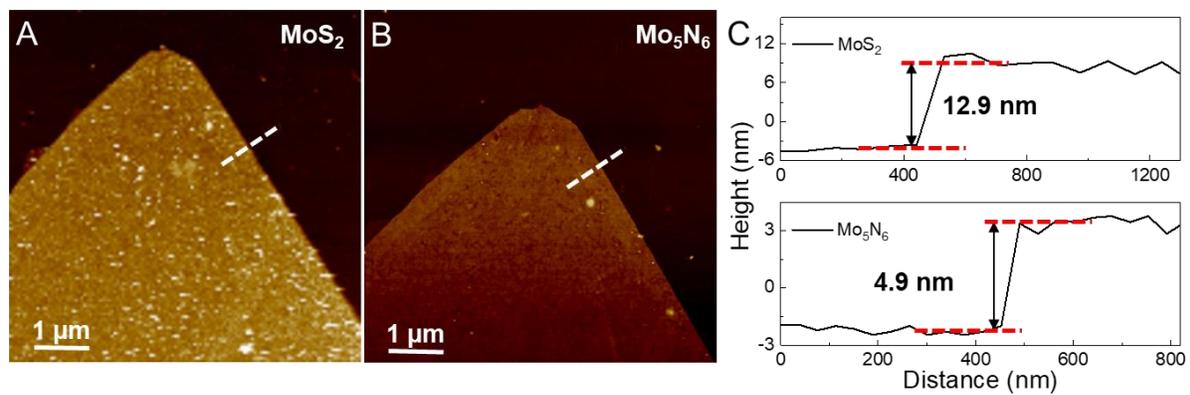

**Fig. S3. AFM images of MoS$_2$ and Mo$_5$N$_6$ flakes in Figure (2). (A)** AFM image of MoS$_2$ flake. **(B)** AFM image of Mo$_5$N$_6$ flake. White dash lines indicate the position where heights are extracted. **(C)** Height profiles of MoS$_2$ and Mo$_5$N$_6$ flakes shown in **(A)** and **(B).**

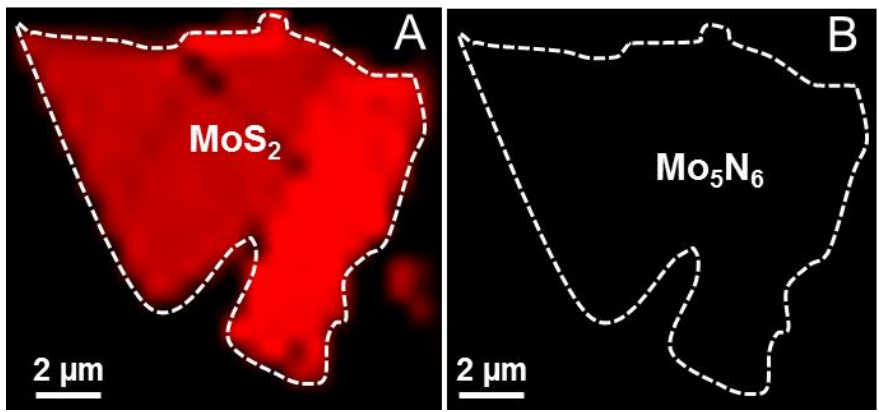

**Fig. S4. Maps of PL intensity at 673 nm of MoS$_2$ (A) and Mo$_5$N$_6$ (B) flakes in Figure 2.** The white dashed lines circle the contour of MoS$_2$ flakes shown in **(A)**.

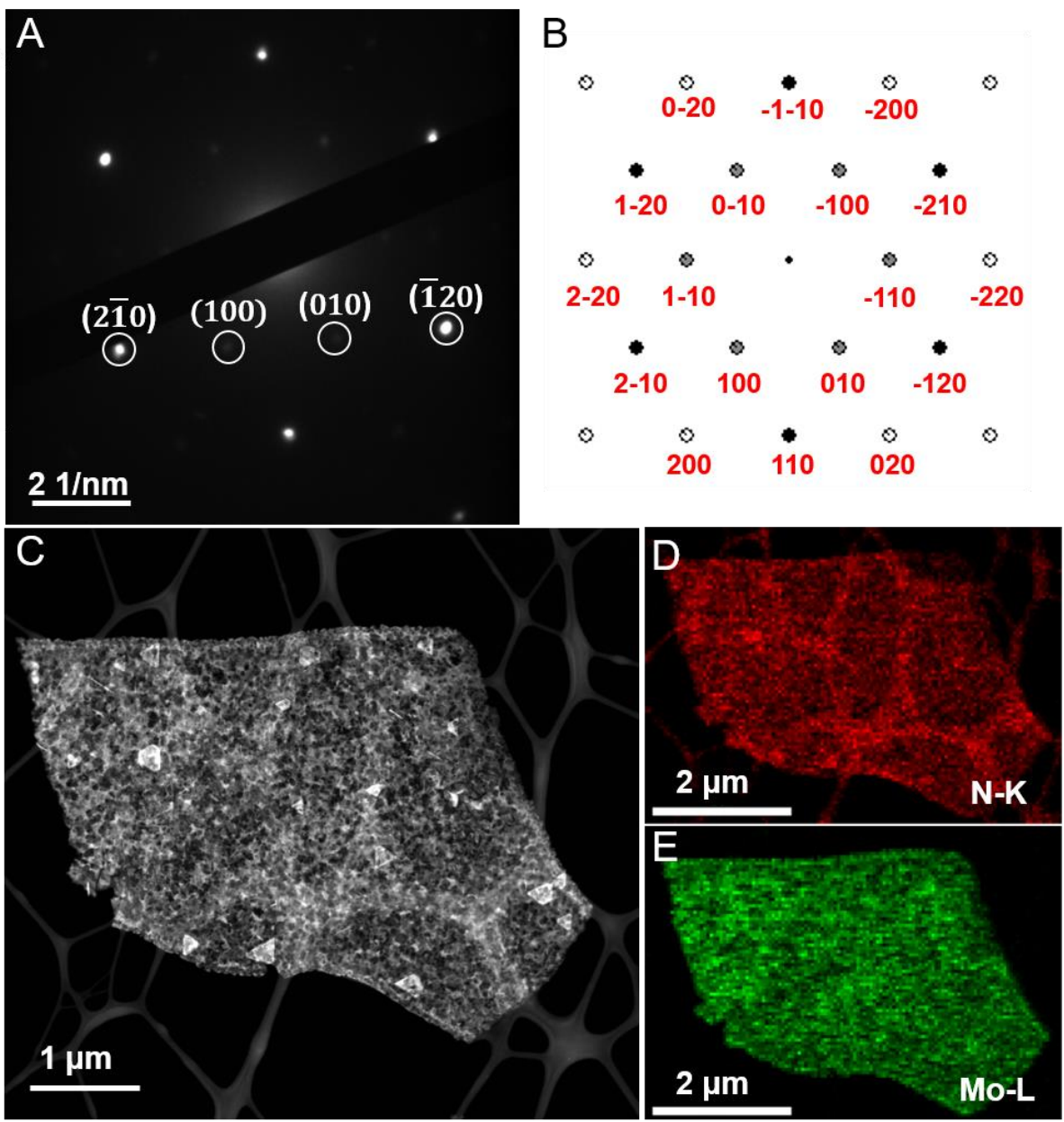

**Fig. S5. Structural and elemental characterizations of Mo$_5$N$_6$.** (**A**) SAED pattern of Mo$_5$N$_6$. (**B**) Simulated SAED pattern of Mo$_5$N$_6$. (**C**) Low magnification dark field STEM image of a thick Mo$_5$N$_6$ flake. (**D** and **E**) EDS mapping of N K peak (**D**) and Mo L peak (**E**) from Mo$_5$N$_6$ flake shown in (**C**).

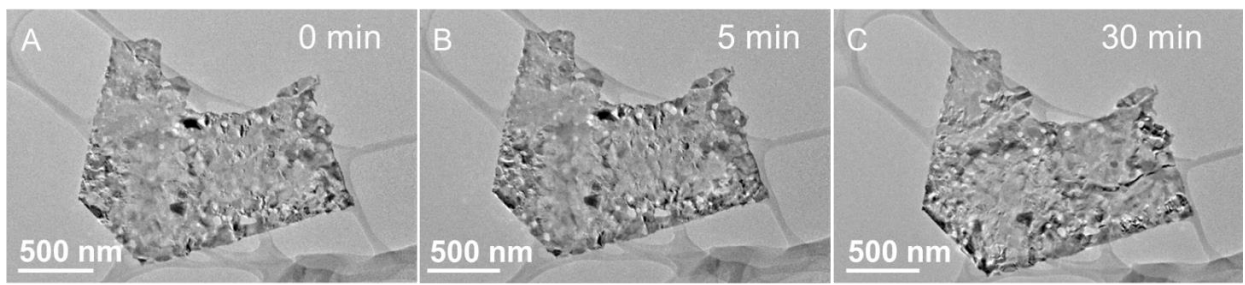

**Fig. S6. TEM images of Mo₅N₆ sample under 200 keV electron beam.** (**A** to **C**) TEM images of Mo₅N₆ after 0 min **(A)**, 5 min **(B)** and 20 min **(C)** exposure under 200 keV electron beam.

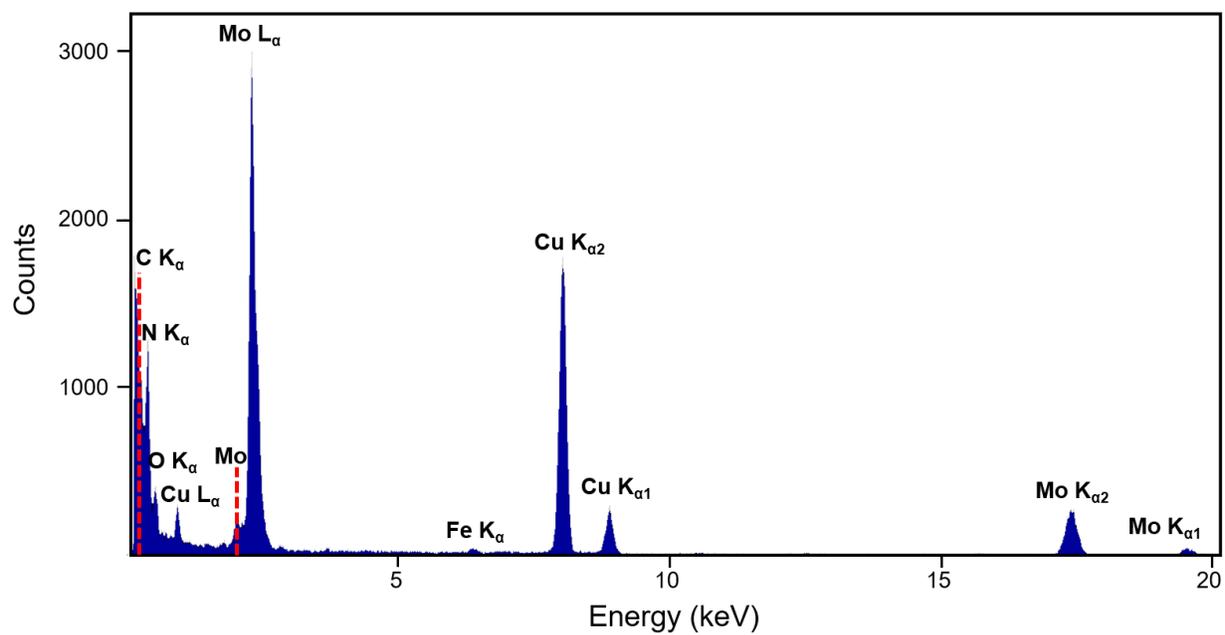

**Fig. S7. EDS spectrum of Mo$_5$N$_6$.** Cu and C peaks are from TEM grid. Fe peaks are from the TEM instrument.

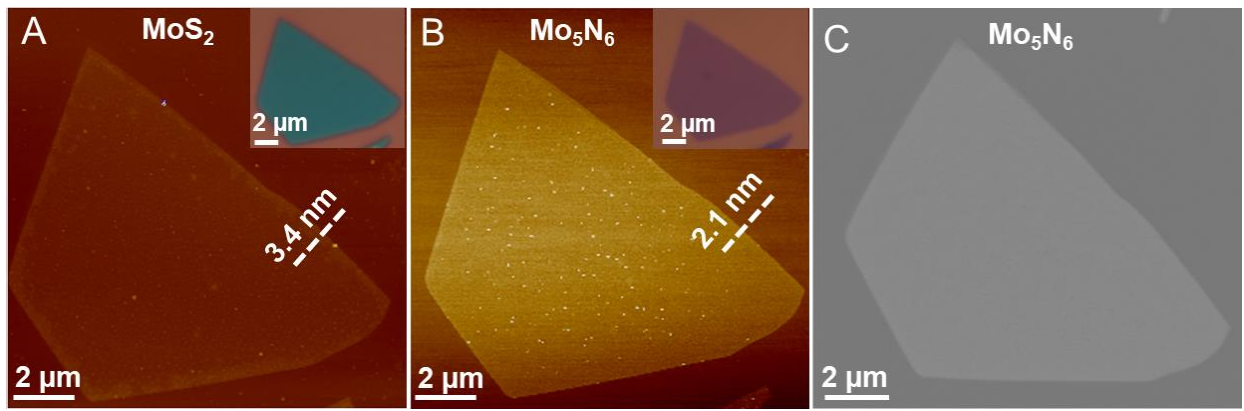

**Fig. S8. Optical, AFM and SEM images of chemical transformation on a 4-layer MoS₂ flake.** (**A**) AFM image of MoS$_2$ flake. Inset: Optical image of MoS$_2$ flake. (**B**) AFM image of Mo$_5$N$_6$ flake. Inset: Optical image of Mo$_5$N$_6$ flake. White dashed lines show the location where thickness of the flakes is measured. (**C**) SEM image of Mo$_5$N$_6$ flake. AFM and SEM images show smooth surface after chemical transformation.

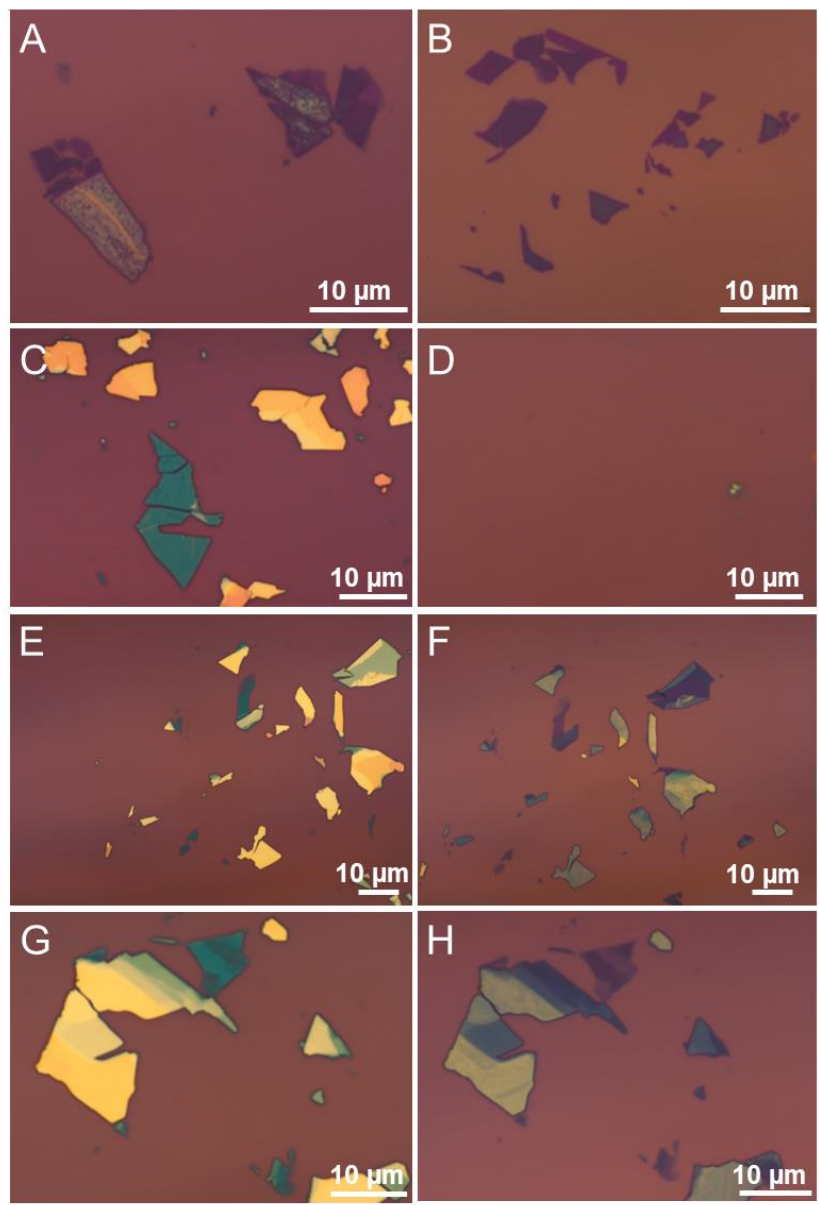

**Fig. S9. Optical images of Mo$_5$N$_6$ flakes prepared from chemical transformations at different conditions.** (**A**) Optical images of Mo$_5$N$_6$ samples converted with 100 mg urea. (**B**) Optical images of Mo$_5$N$_6$ samples converted with 500 mg urea. (**C** and **D**) Optical images of samples before (**C**) and after (**D**) conversion without urea. (**E** and **F**) Optical images of samples before (**E**) and after conversion (**F**) at 800 °C with 500 mg urea. (**G** and **H**) Optical images of samples before (**G**) and after conversion (**H**) at 750 °C with 500 mg urea.

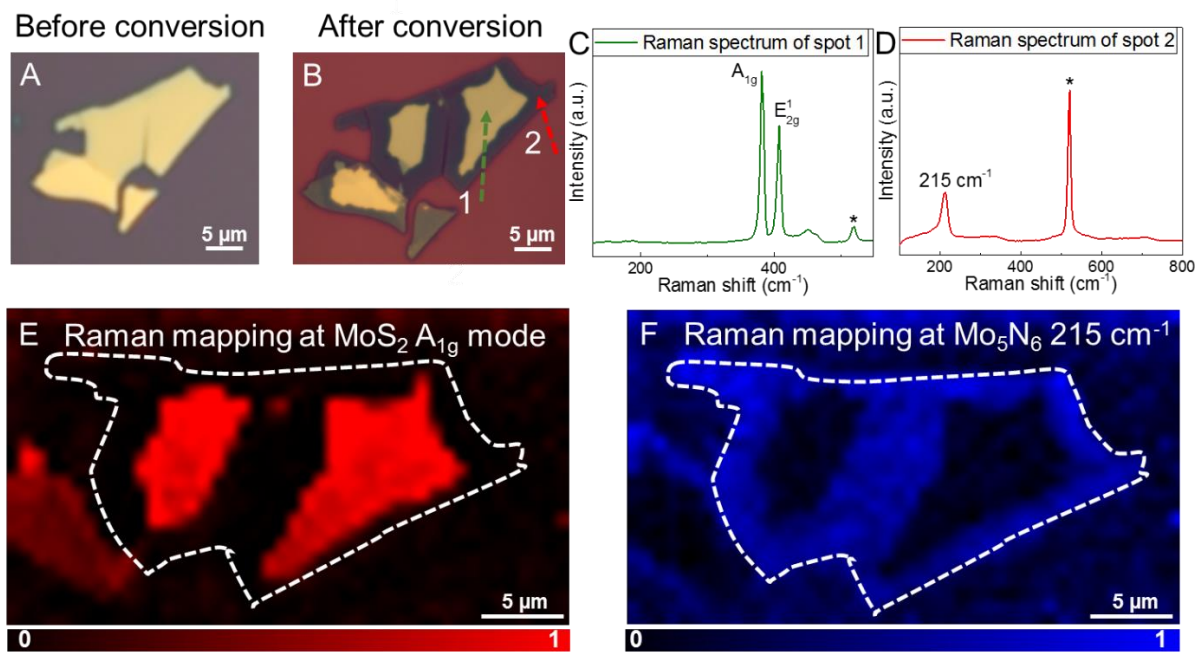

**Fig. S10. Optical images and Raman spectra of a partially converted flake at 700 °C.** (**A** and **B**) Optical images of flake before conversion-MoS$_2$ (**A**) and after conversion-Mo$_5$N$_6$ (**B**). (**C** and **D**) Raman spectra of flake after partial conversion at spot 1 (**C**) and spot 2 (**D**). Peak labeled with * is from SiO$_2$/Si substrate. (**E** and **F**) Raman mappings of the flake at MoS$_2$ A$_{1g}$ mode (**E**) and Mo$_5$N$_6$ 214 cm$^{-1}$(**F**). The white dashed lines circle the contour of MoS$_2$ flakes shown in (**B**).

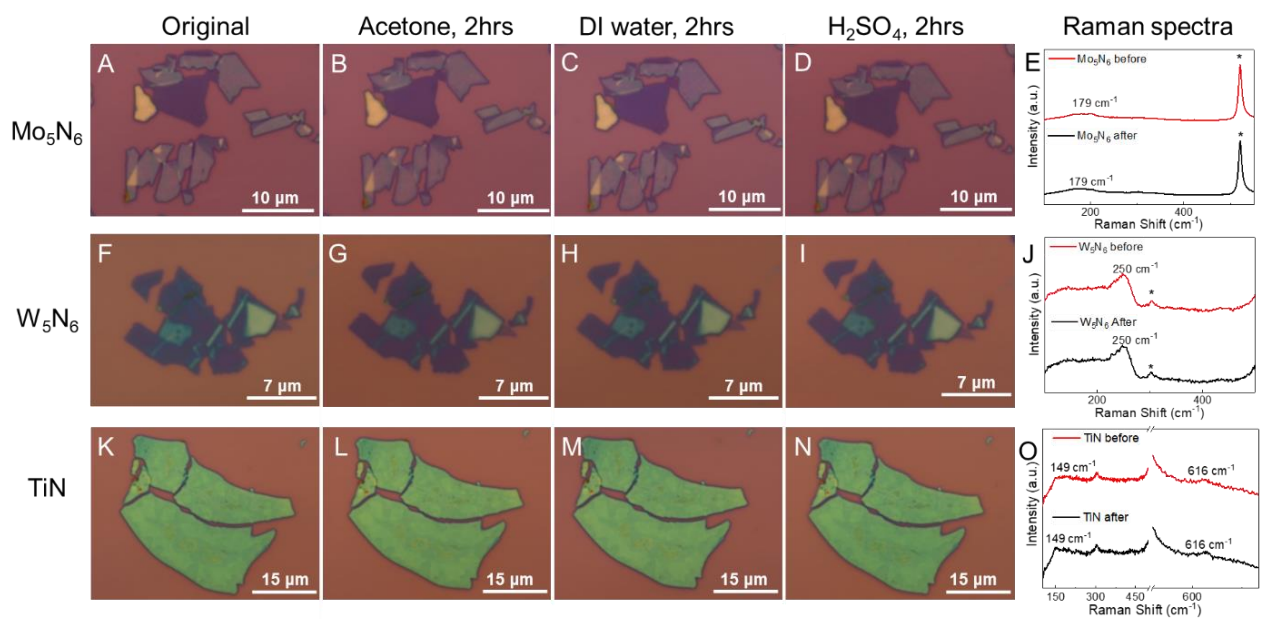

**Fig.S11. Stability test of Mo₅N₆, W₅N₆ and TiN.** (**A to D**) Optical images of Mo₅N₆ flakes before any treatment (**A**), after two-hour immersion in acetone (**B**), after two-hour immersion in deionized water (**C**), and after two-hour immersion in 1mol/L $H_2SO_4$ solution (**D**). (**E**) Raman spectra of a typical Mo₅N₆ sample before and after the stability tests. (**F to I**) Optical images of W₅N₆ flakes before any treatment (**F**), after two-hour immersion in acetone (**G**), after two-hour immersion in deionized water (**H**), and after two-hour immersion in 1mol/L $H_2SO_4$ solution (**I**). (**J**) Raman spectra of a typical W₅N₆ sample before and after the stability tests. (**K to O**) Optical images of TiN flakes before any treatment (**K**), after two-hour immersion in acetone (**L**), after two-hour immersion in deionized water (**M**), and after two-hour immersion in 1mol/L $H_2SO_4$ solution (**N**). (**O**) Raman spectra of a typical TiN sample before and after the stability tests.

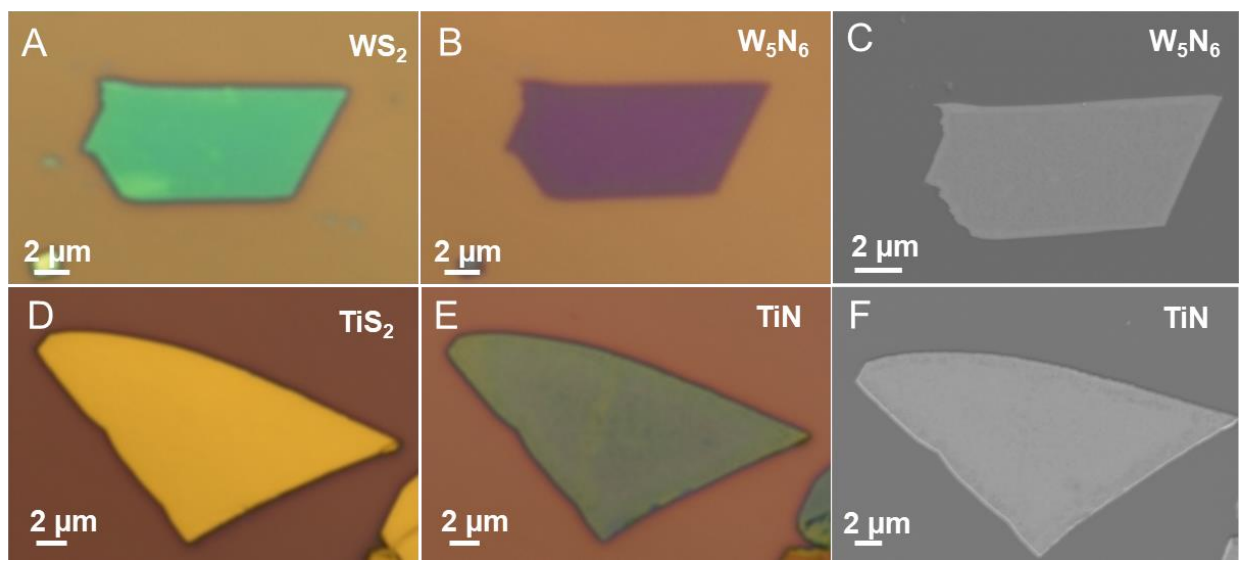

**Fig. S12. Optical and SEM images of WS₂, W₅N₆, TiS₂, and TiN flakes.** (**A**) Optical image of a WS₂ flake. (**B**) Optical image of a W₅N₆ flake. (**C**) SEM image of a W₅N₆ flake. (**D**) Optical image of a TiS₂ flake. (**E**) Optical image of a Mo₅N₆ flake. (**F**) SEM image of a TiN flake.

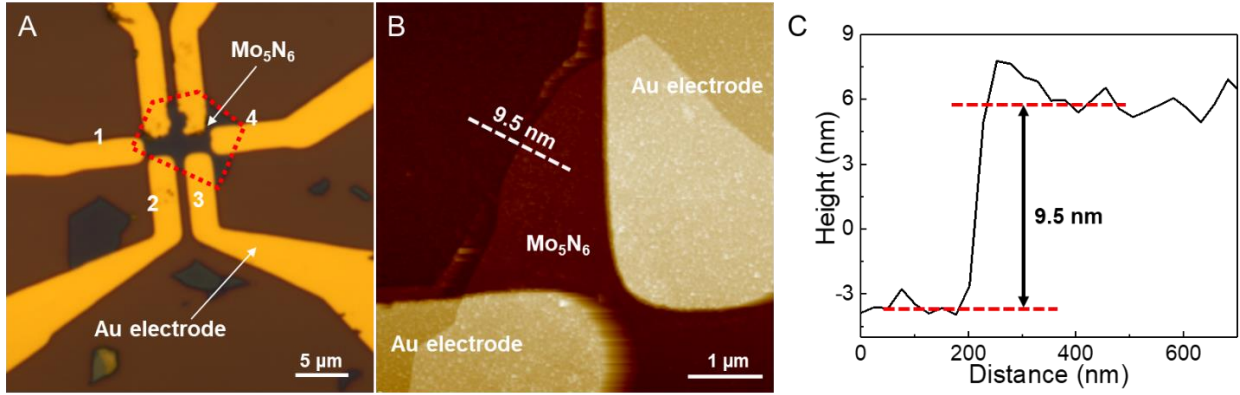

**Fig. S13. Optical and AFM images of Mo₅N₆ transport device.** (**A**) Optical image of a 6-terminal transport device of Mo$_5$N$_6$. Area circled by the red curve is the Mo$_5$N$_6$ flake that is used for device. Au electrodes from 1 to 4 are labelled and used for 4-probe electrical measurements. (**B**) AFM image of the Mo$_5$N$_6$ transport device. The thickness of the flake is ~ 9.5 nm. White dashed line indicated the location where thickness was extracted. (**C**) Height profile of Mo$_5$N$_6$ flake in the transport device.

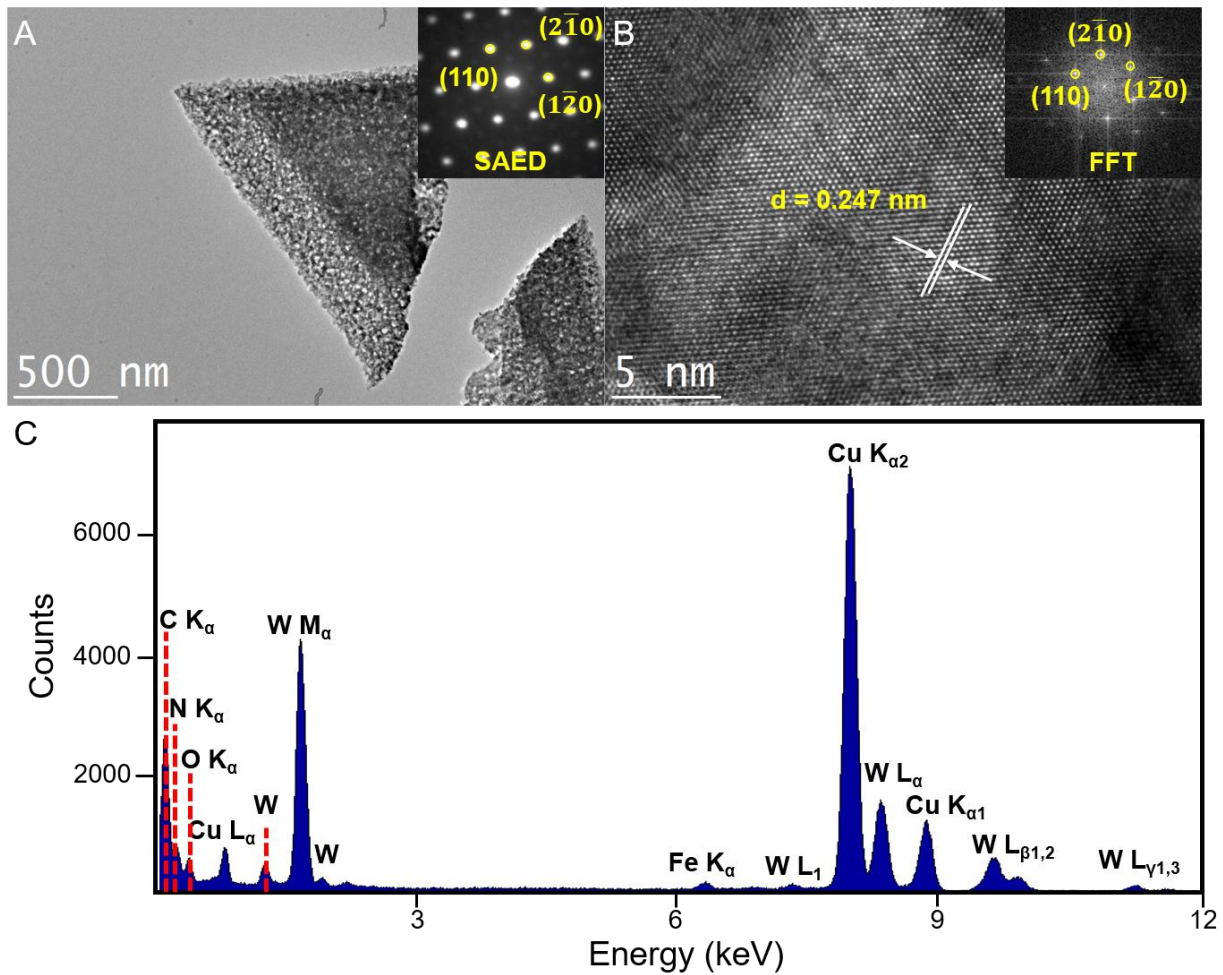

**Fig. S14. TEM and EDS characterizations of $W_5N_6$ converted from $WS_2$.** (**A**) Low magnification TEM image of $W_5N_6$. Inset: Selected area electron diffraction (SAED). (**B**) High resolution TEM image of $W_5N_6$. Inset: FFT of the image. (**C**) EDS spectrum of $W_5N_6$ flake. Cu peaks are from TEM grid.

To determine the phase and crystal structure of the flake, we performed transmission electron microscope (TEM) measurement on an as-prepared sample after transferred onto a TEM grid. The low magnification TEM image in fig. S12A shows the flake on a TEM grid, the rough surface is probably caused by KOH etching during the transfer process. The strong selected-area electron diffraction (SAED) pattern (fig. S12A inset) indicates the high crystallinity of the crystal. The hexagonal pattern is consistent with the crystal structure of the $W_5N_6$ (*48*). High-resolution TEM (HRTEM) image is shown in fig. S12B, where no obvious defect is found in the sample. Additionally, the distance of 0.247 nm between the highlighted lattice planes is consistent with the d-spacing of (110) planes in $W_5N_6$ crystal, which is labeled in the fast Fourier transformation (FFT) pattern in the inset. Other two planes ($2\bar{1}0$) and ($1\bar{2}0$) are also labeled in the FFT pattern. EDS spectrum (fig. S12C) shows the presence of W and N elements and no peaks of S element is observed.